\documentclass[conference]{IEEEtran}
\usepackage[T1]{fontenc}
\usepackage{graphicx}
\usepackage{xcolor}
\usepackage{amsmath,amssymb,amsfonts}
\usepackage{subcaption}
\usepackage{hyperref}
\usepackage{tabularx}
\usepackage{booktabs}
\usepackage{makecell}
\usepackage{multirow}
\usepackage{pdfpages}

\graphicspath{{figs/}}

\newcommand{\mycolor}{black}
\newcommand{\mycolorrev}{black}

\usepackage{ifthen}
\newboolean{showcomments}
\setboolean{showcomments}{false}          
\ifthenelse{\boolean{showcomments}}
  {\newcommand{\nb}[2]{
    \fcolorbox{gray}{yellow}{\bfseries\scriptsize#1}
    {\scriptsize$\blacktriangleright${#2}$\blacktriangleleft$}
   } 
  }
  {\newcommand{\nb}[2]{}}

\newcommand{\adem}[1]{\nb{Adem}{#1}}

\title{Towards Modeling Human-Agentic Collaborative Workflows: A BPMN Extension}

\author{
    \IEEEauthorblockN{Adem Ait}
    \IEEEauthorblockA{\textit{University of Luxembourg}\\
    Esch-sur-Alzette, Luxembourg\\
    adem.ait@uni.lu}
    \and
    \IEEEauthorblockN{Javier Luis C\'anovas Izquierdo}
    \IEEEauthorblockA{\textit{IN3 - UOC} \\
    Barcelona, Spain \\
    jcanovasi@uoc.edu}
    \and
    \IEEEauthorblockN{Jordi Cabot}
    \IEEEauthorblockA{\textit{University of Luxembourg} \\
    \textit{Luxembourg Institute of Science and Technology}\\
    Esch-sur-Alzette, Luxembourg \\
    jordi.cabot@list.lu}
}

\begin{document}

\maketitle

\begin{abstract}
  Large Language Models (LLMs) have facilitated the definition of autonomous intelligent agents.
  Such agents have already demonstrated their potential in solving complex tasks in different domains. 
  And they can further increase their performance when collaborating with other agents in a multi-agent system.
  However, the orchestration and coordination of these agents is still challenging, especially when they need to interact with humans as part of human-agentic collaborative workflows. 
  These kinds of workflows need to be precisely specified so that it is clear whose responsible for each task, what strategies agents can follow to complete individual tasks or how decisions will be taken when different alternatives are proposed, among others. 
  Current business process modeling languages fall short when it comes to specifying these new mixed collaborative scenarios.
  In this exploratory paper, we extend a well-known process modeling language (i.e., BPMN) to enable the definition of this new type of workflow.
  Our extension covers both the formalization of the new modeling concepts required and the proposal of a BPMN-like graphical notation to facilitate the definition of these workflows. 
  Our extension has been implemented and is available as an open-source human-agentic workflow modeling editor on GitHub.  
\end{abstract}

\section{Introduction}
\label{sec:introduction}
In our current information-rich society, the integration of agents, especially agents powered by Large Language Models (LLMs), is becoming more and more important to quickly perform many tasks~\cite{xi2025rise}. 
Agents can interact with the environment, make their own decisions, and learn from the received feedback. 
Moreover, often, agents do not work in isolation but as part of Multi-Agent Systems (MAS) where agents cooperate (or compete) to achieve a common goal~\cite{DBLP:conf/ijcai/GuoCWCPCW024}. 
This collaborative process is known as an agentic system. 
These systems have already demonstrated their superior performance against single-agent solutions~\cite{DBLP:conf/ijcai/GuoCWCPCW024}.

While agentic systems are performant in many tasks, complex scenarios require the participation of humans~\cite{DBLP:journals/corr/abs-2308-08155}.
Therefore, there is a need to precisely define this collaboration and how each participant interacts with each other. 
Unfortunately, we argue that current process modeling languages, such as BPMN, lack the modeling constructs to specify this collaboration between humans and agentic systems as new primitives to define the confidence of the agents, the strategies they can use to perform a task, or the process to reach a decision.
Furthermore, frameworks targeting the implementation of agentic workflows, such as \textsc{LangGraph}\footnote{\url{https://www.langchain.com/langgraph}}, minimize the participation of humans in the process and therefore are not expressive enough to model the human-agent interaction beyond very simple cases.

The goal of this paper is to enable the precise definition of human-agentic workflows. 
To this aim, we study how the Business Process Model and Notation (BPMN), one of the most well-known modeling languages for workflows, could be used to represent this new type of workflow, and then, based on the identified limitations, we propose a BPMN extension to enable their definition in BPMN. 
\textcolor{\mycolor}{Note that our approach can serve as a blueprint for extending other workflow languages, as the conceptual elements we identify are largely notation-independent.} 
This extension has been implemented in an open source modeling tool available on \textsc{GitHub}. 

The rest of the paper is structured as follows.
Section~\ref{sec:background} provides the background and a running example.
Section~\ref{sec:motivation} shows the mapping of agentic concepts to BPMN, while Section~\ref{sec:extension} presents the extension to the BPMN to overcome the limitations found.
Section~\ref{sec:bpmn-x} and ~\ref{sec:toolSupport} provide the extension definition and the proof of concept, respectively.
Section~\ref{sec:related} presents the related work.
Finally, Section~\ref{sec:conclusion} concludes the paper and presents the roadmap.

\section{Background \& Running Example}
\label{sec:background}
In this section, we briefly describe BPMN, the language we aim to extend; and present the main concepts of agentic systems. 
We end the section with a running example.

\begin{figure*}[t]
    \centering
    \begin{subfigure}[t]{.85\textwidth}
        \centering
        \includegraphics[width=\textwidth]{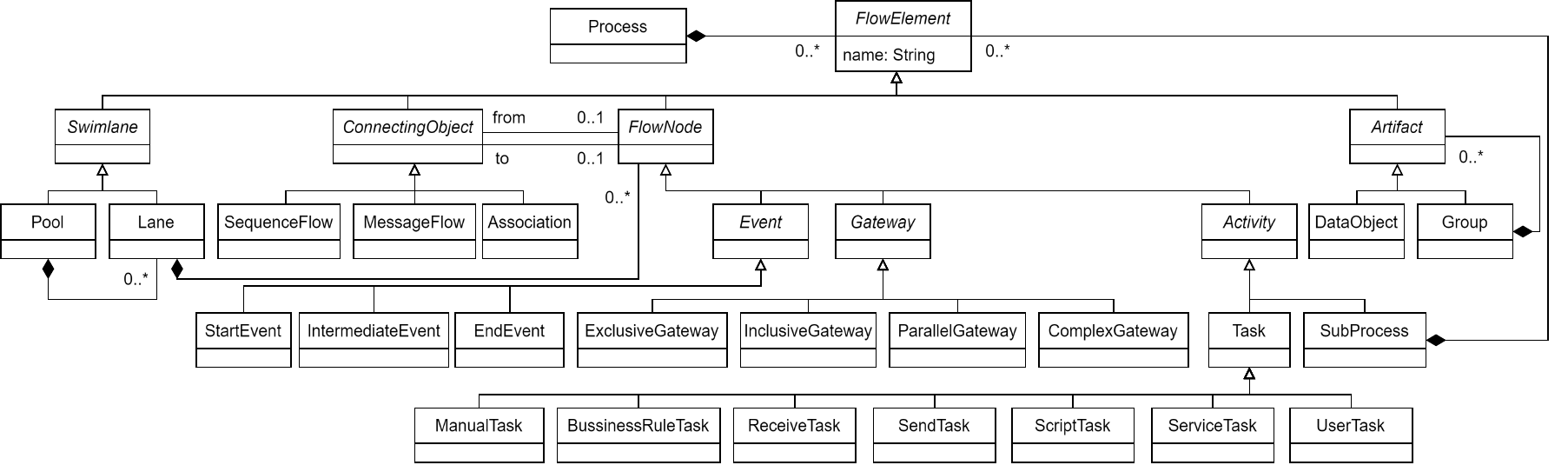}
        \caption{}
        \label{fig:bpmn:metamodel}
    \end{subfigure}
    \hfill
    \begin{subfigure}[t]{.55\textwidth}
        \centering
        \includegraphics[width=\textwidth]{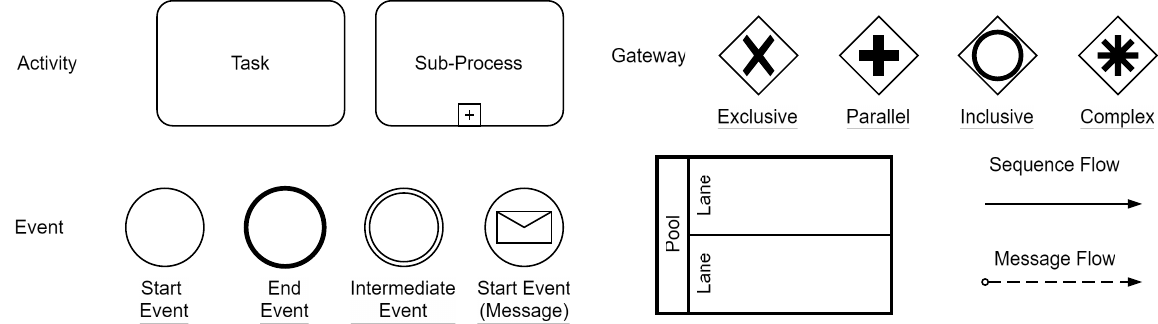}
        \caption{}
        \label{fig:bpmn:main-elements}
    \end{subfigure}
    \caption{Simplified (a) metamodel of BPMN and the corresponding (b) notation.}
    \label{fig:bpmn}
\end{figure*}

\subsection{BPMN}
\label{sec:background:bpmn}

Business Process Management (BPM) studies how work is performed in an organization to ensure consistent outcomes and to take advantage of improvement opportunities~\cite{DBLP:books/daglib/0031128}. 
BPM is about managing entire chains of events, activities, and decisions (called business processes) that ultimately add value to the organization and its customers. 
Thus, the formal representation of such process facilitates the orchestration, monitoring, and improvement of an organization’s workflow.

BPMN~\cite{BPMN_Specification_2014} has become the de-facto standard for business processes diagrams.
It is defined by the Object Management Group (OMG) and specified as ISO standard (ISO/IEC 19510:2013).
BPMN provides a metamodel and a notation to define and visualize business process models. 
It provides an extension mechanism to allow modeling domain-specific elements not included in the specification.
In particular, BPMN 2.0 extension mechanism enables the approach of extension by addition, which consists of attaching new domain-specific elements to the predefined elements of the language.

Figure~\ref{fig:bpmn:metamodel} shows a simplified metamodel of BPMN.
The different \textit{FlowElements} compose the \textit{Process}.
The \textit{Swimlane} is used to organize the \textit{Process}, which can be either \textit{Pool}, or \textit{Lane}. 
\textit{Pools} can be formed by several \textit{Lanes}. 
\textit{FlowObject} can be: 
(1) \textit{Events}, which are used as triggers (e.g., timers); 
(2) \textit{Activities}, the work unit, which can be atomic (i.e., \textit{Tasks}) and non-atomic (i.e., \textit{Sub-Process}); 
and (3) \textit{Gateways} for controlling the flow.
\textit{ConnectingObject} is used to specify the order of \textit{FlowElements} which is divided into: 
(1) \textit{SequenceFlow}, to establish the flow within pools, which can also be conditional (i.e., the flow will continue if a condition is fulfilled); 
(2) \textit{MessageFlow}, to communicate between pools; 
and (3) \textit{Association}, used to connect user-defined text (an \textit{Annotation}) with \textit{Flow Nodes}.
\textit{Artifact} describes the \textit{DataObject} shared in the process and \textit{Group}, a visual mechanism to group elements of a diagram informally.
Figure~\ref{fig:bpmn:main-elements} shows the notation of the main elements described.

\subsection{Agentic systems}
\label{sec:background:llm-agents}
LLMs excel at multiple tasks~\cite{DBLP:journals/tmlr/WeiTBRZBYBZMCHVLDF22}, but creating specialized LLMs instances to target specific tasks has shown promising results.
These specialized LLMs are typically known as agents~\cite{xi2025rise}. 
LLM-based agents use the potential of LLMs, facilitating sophisticated interactions, decision-making, tool-use capabilities, and in-context learning through memory~\cite{DBLP:conf/ijcai/GuoCWCPCW024}.

However, LLMs exhibit non-deterministic behavior. 
To alleviate this, reflection strategies have been proposed to refine their answers~\cite{DBLP:conf/nips/ShinnCGNY23}, namely: 
(1) \textbf{self-reflection}, where agents generate feedback on the plan and reasoning process to refine themselves; 
(2) \textbf{cross-reflection}, where the feedback is provided by other agents; 
and (3) \textbf{human-reflection}, where humans provide the feedback. 

To tackle more complex problems, a common approach is to increase the number of agents, forming what is known as a LLM-based MAS, or, more recently, rebranded as agentic systems, to emphasize the cooperation of the agents in the MAS which has already been proven to outperform single-agent solutions~\cite{DBLP:conf/ijcai/GuoCWCPCW024}.
A key aspect in agentic systems is how agents work collaboratively to solve tasks, leveraging their interactions with the environment or other agents.
The works by Guo et al.~\cite{DBLP:conf/ijcai/GuoCWCPCW024} and Liu et al.~\cite{DBLP:journals/jss/LiuLLZZXHW25} introduce the first attempts at characterizing how agentic systems work.

Agentic systems can adhere to different types of cooperation patterns. The core ones are: 
(1) \textbf{voting-based}, where agents independently propose alternative solutions and reach consensus by voting; 
(2) \textbf{role-based}, where each agent, or group of agents, has assigned a role, thus making the decision according to such roles; 
and (3) \textbf{debate-based}, where agents submit and receive feedback to adjust the thoughts until a consensus is reached.
Furthermore, an additional scenario is \textbf{competition-based} collaboration, where agents, instead of cooperating, compete and the fastest (or the most reliable output) is selected.

All these agentic aspects will need to be part of the process modeling language if we want to model in detail human-agentic collaborations as, for instance, in the scenario we use as running example. 

\subsection{Running Example}
\label{sec:running-example}

To illustrate our proposal, we will use a running example based on a simple resolution process for bug reports in a software project.
The example process comprises five participants, two humans and three agents.

The humans are a user, who reports the bug; and a maintainer, who reviews the final change proposal and resolves the bug. 
The agents are responsible for solving the bug by implementing change proposals and deciding together the best option.
There are three agents, one is used as a reviewer and the other two are specialized coding agents.
Each agent comes with a level of uncertainty regarding the quality of all its actions. 
This value could be derived from the underlying LLM and/or the agent setup. 

Once the reviewer agent validates the bug definition, using a reflective strategy to double-check on the first assessment, the two coding agents are in charge of proposing a solution to the bug. They both work independently in parallel and, following a role-based cooperation strategy, is the agent with the reviewer role who has the final decision, also considering the uncertainty of each coding agent in case of discrepancies.

\section{Using BPMN to Model Human-Agentic Workflows}
\label{sec:motivation}
In this section, we study the current support of BPMN to model human-agentic workflows.
The characteristics of agentic systems we aim at covering are the ones that concern the reflection and cooperation, while also addressing their non-deterministic behavior.
Table~\ref{tab:mapping-agents-to-bpmn} shows a possible mapping of BPMN elements to model human-agentic workflows. 
Note that this mapping is partial, as the current support of BPMN for this type of workflow is very limited, as we show in this section.

\begin{table}[t]
    \centering
    \scriptsize
    \caption{Partial mapping of human-agentic workflow concepts to BPMN elements.}
    \label{tab:mapping-agents-to-bpmn}
    \begin{tabularx}{.49\textwidth}{llX}
    \textsc{\makecell[l]{human-agentic\\Workflow Concept}} & \textsc{Variety} & \textsc{BPMN Element}     \\
    \toprule
    \multirow{2}{*}{Agent}                          & Single-agent                & Pool or Lane                     \\
                                                    & Multi-agent                 & Multi-instance pool \\
    \midrule
    \multirow{3}{*}{Reflection}                     & Self-reflection             & Extra loop activity                    \\
                                                    & Cross-reflection            & Loop with gateways and activities \\
                                                    & Human-reflection            & Loop with gateways and activities \\
    \midrule
    Agent Collaboration                             & All                         & Group or message flow          \\
    \midrule
    \makecell[l]{Merging\\ Collaboration Efforts}   & All                         & Complex gateway or message flow \\
    \bottomrule
    \end{tabularx}
\end{table}

Agents could be represented as pools or lanes, depending on the process context.
When agents are part of a project process, lanes could be used, but if they represent external contributors, pools could be used instead.
Sets of agents could be represented as multi-instance pools.
Information about the non-deterministic behavior of the agents could only be represented via text annotations.

Reflection strategies cannot be represented in standard BPMN. 
We could simulate them using additional activities and gateways.
Self-reflection could be represented as a loop activity that follows the activity to be refined.
For cross- and human-reflection, the feedback loop could be represented as several gateways and activities to keep refining the answer until the desired output is obtained.

To model the collaboration between agents, we consider two scenarios, depending on whether agents are located in lanes of the same or different pool/s (i.e., collaboration diagram).
When located in the same pool, gateways could be used to route the flow towards multiple lanes, and groups could be utilized to set the collaboration strategy.
If located in different pools, message flows could be used, but text annotations should be added to describe the strategy.
For merging collaborations, in the first scenario we could use a complex gateway and specify the desired merging strategy as an annotation, while in the second scenario we could specify the strategy as an annotation in the incoming message flow.
Note the need to use text annotations to describe the behavior. 
\textcolor{\mycolorrev}{BPMN also offers additional collaboration mechanisms such as choreographies (formalized coordination of interactions between participants), but these still require supplementary descriptions for agent-specific behaviors since their focus is on the exchange of information rather than the orchestration of their work.}

\begin{figure*}[t]
    \centering
    \includegraphics[width=.8\textwidth]{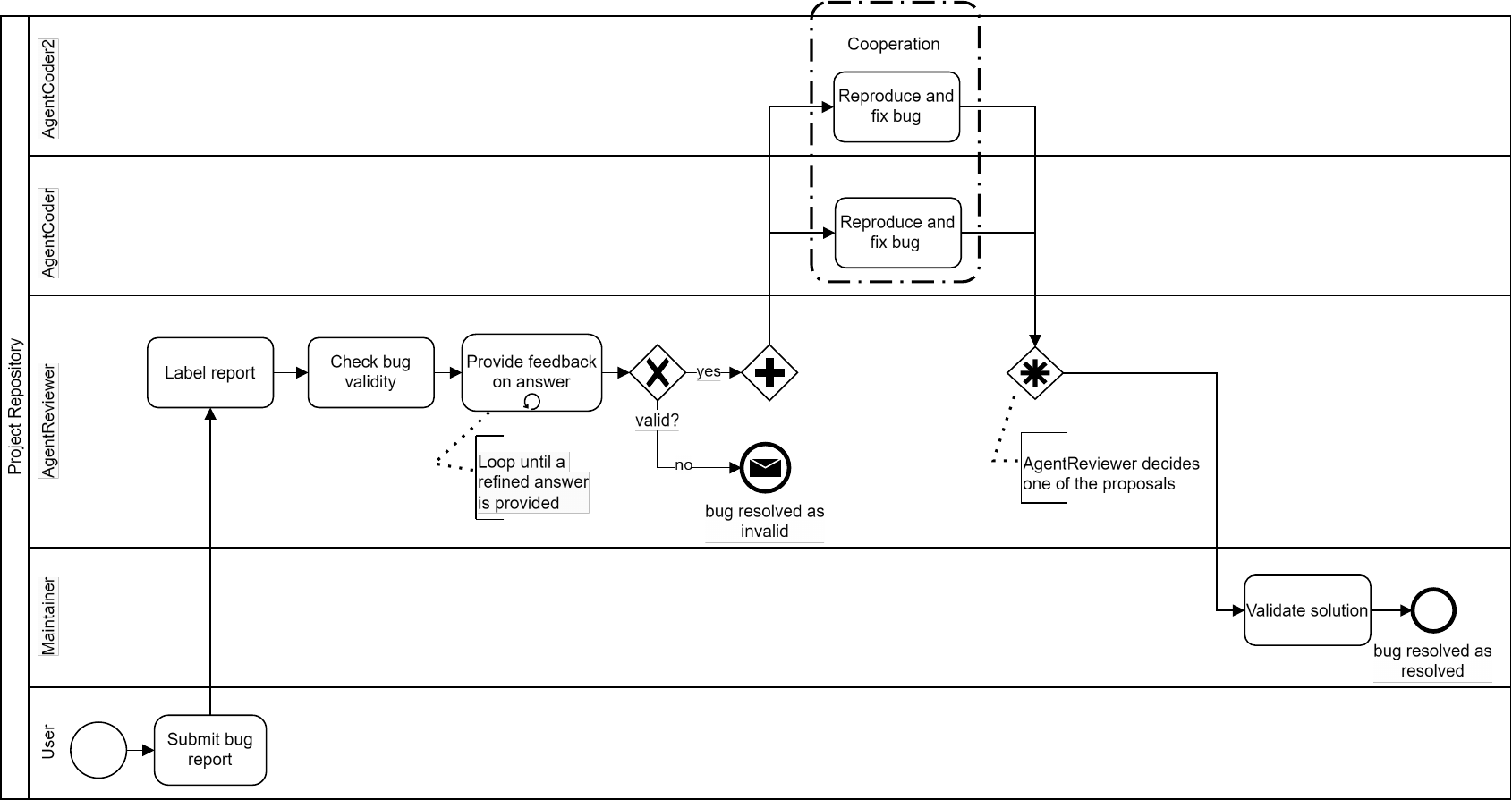}
    \caption{Running example with standard BPMN.}
    \label{fig:running-example-bpmn-basic}
\end{figure*}

Figure~\ref{fig:running-example-bpmn-basic} shows the BPMN model for the running example.
As can be seen, we used five lanes to represent the different actors.
The \textit{User} and \textit{Maintainer} are the humans reporting the bug and validating the final proposed solution, respectively.
The \textit{AgentReviewer} is the agent designated to solve the bug, while \textit{AgentCoder} and \textit{AgentCoder2} represent specialized coding agents that will help with coding tasks.
To apply self-reflection on the \textit{Check bug validity} activity, we define an extra loop activity (see \textit{Provide feedback on answer}) with a loop condition defined in natural language to not stop until a refined answer is provided.
Once we obtain a reliable answer, we proceed to fix the bug if the report is valid.
To illustrate the cooperation, we use a group (see \textit{cooperation}).
Then, as a merging gateway, we use a complex gateway to define our own condition through an annotation, which is decided by \textit{AgentReviewer}.
Thus, only the selected solution will be the one delivered to the next activity.

Representing human-agentic workflows in BPMN is challenging, as the standard BPMN does not provide specific elements to model agents or their interactions.
In the running example, we cannot set uncertainty to determine the agents' reliability or identify lanes or pools as agents.
We can partially represent the collaboration and reflection strategies.
We rely on natural language to describe the collaboration and reflection strategies, which can lead to ambiguity and misinterpretation.
Furthermore, the intricate process of representing reflection can hinder the readability and understandability of the model.

\textcolor{\mycolor}{BPMN was chosen for several key reasons: (1) widespread adoption, BPMN is the \emph{de facto} standard for business process modeling with broad industrial and academic support; (2) expressivity, BPMN already provides comprehensive constructs for modeling workflows involving human participants; (3) extensibility, BPMN offers an extension mechanism that enables domain-specific additions while maintaining compliance with the standard; (4) execution capabilities, BPMN models can be directly executed by business process engines, making the transition from conceptual models to implementation easier; and (5) integration with existing processes, many organizations already use BPMN, making it easier to incorporate agent-based elements into existing business processes.}\adem{Too extensive? Maybe it is also valuable to add some references in some statements (e.g., widespread use or expressivity).}

\textcolor{\mycolor}{Therefore, despite the identified limitations, we believe that BPMN still provides a good starting point compared to alternative modeling languages. 
While languages like UML Activity Diagrams or YAWL could potentially be extended in similar ways, BPMN's widespread usage in industries means that practitioners are already familiar with its core concepts, reducing the learning curve for our extended notation.}

In the next section, we propose an extension to BPMN that introduces formalized syntax for defining collaboration and reflection strategies, rules for merging collaborative outcomes, and agents' profiling (i.e., role and trustworthiness).

\section{Extending BPMN to Model Human-Agentic Workflows}
\label{sec:extension}
To address the limitations identified for modeling human-agentic workflow concepts in BPMN, we propose an extension to the BPMN specification.
In particular, we propose to extend the following elements: lane, task, message flow, and parallel and inclusive gateways.
In the following, we describe each extension point, showing the domain models of the extended BPMN elements.
When illustrating the extension, classes of the BPMN 2.0 metamodel will be highlighted in gray.
Note that, without loss of generality, we describe our extension as a BPMN extension, but a similar approach could have been used to extend other process modeling languages, as they offer a largely overlapping set of concepts.

\subsection{Agent Profiling}
\label{sec:extension:agenticLane}

\begin{figure}[t]
    \centering
    \includegraphics[width=.6\columnwidth]{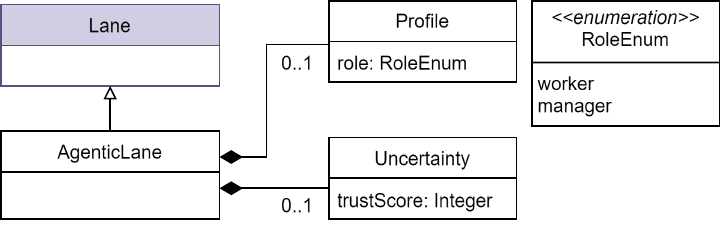}
    \caption{Domain model of the \textit{AgenticLane}.}
    \label{fig:model-agentic-lane}
\end{figure}

Agent profiling requires modeling (1) the role of the agents in collaboration scenarios, as it is how an LLM is initialized as an agent; and (2) the reliability of the agent's output, or trust score, to address the non-deterministic nature of agents behavior.
As shown in Section~\ref{sec:motivation}, in standard BPMN, representing agents as lanes or pools overlooks these aspects: (1) when having multiple lanes, each representing specialized agents, only lane names can be specified; and (2) there is no way to specify uncertainty for agent's output. 

Figure~\ref{fig:model-agentic-lane} shows the domain model of the extension to address agent profiling.
We differentiate between regular participants and \textit{agentic} participants (see \emph{Lane} and \emph{AgenticLane}).
The \textit{Pool} element is the graphical representation of a participant.
However, we decided to extend the \textit{Lane} class rather than the \textit{Pool} class to allow setting a profile for each agent within a pool.
This way, a group of agents can be represented as a pool, where each agent can have different trust scores since they would be represented as lanes of the pool.

To represent the role and the trust score, we define the attributes in the \textit{Profile} and \textit{Uncertainty} classes, respectively.
The role value distinguishes between manager and workers, but the corresponding enumeration could be extended to fulfill further roles (e.g., coder).
When set to manager, the agent represented by the lane is the one in charge of selecting the valid output in role-based or debate-based cooperation.
The trust score parameter is a percentage value (i.e., 0-100) of the trustworthiness of a particular agent. 

\subsection{Agent Reflection}
\label{sec:extension:agenticTask}

\begin{figure}[t]
    \centering
    \includegraphics[width=.6\columnwidth]{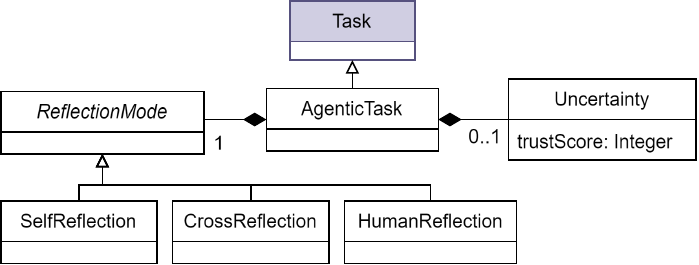}
    \caption{Domain model of the \textit{AgenticTask}.}
    \label{fig:model-agentic-task}
\end{figure}

Reflection is key in agentic systems, enabling agents to evaluate their actions and adapt their behavior accordingly.
Although standard BPMN can model loops and decision points, it lacks the constructs to formally define and enforce self-reflection, cross-reflection, or human-reflection processes. 

Figure~\ref{fig:model-agentic-task} shows the domain model of the extension to address agent reflection. 
We extend the \textit{Task} BPMN element, defining the \textit{AgenticTask}, which can be associated to one of the three reflection strategies.
We model the identified reflection strategies from the literature into three classes (see \textit{SelfReflection}, \textit{CrossReflection}, and \textit{HumanReflection} classes), which are subclasses of the \textit{ReflectionMode} class.
Depending on the degree of reliability and the resources available, one could define the reflection as human-reflection, to avoid undesired outputs.
However, if one wants a completely automated task, but also ensure some degree of reliability, one could use self-reflection or cross-reflection strategies, where the latter would be more expensive, since it requires instances of other agents, but it might provide better results.

Furthermore, tasks can have attached a trustworthiness score (see \emph{Uncertainty}), used to represent the reliability of the task output.
This trust score can be further used in the workflow to decide the next steps.

\subsection{Agent Collaboration}
\label{sec:extension:agentCollaboration}

\begin{figure}[t]
    \centering
    \includegraphics[width=\columnwidth]{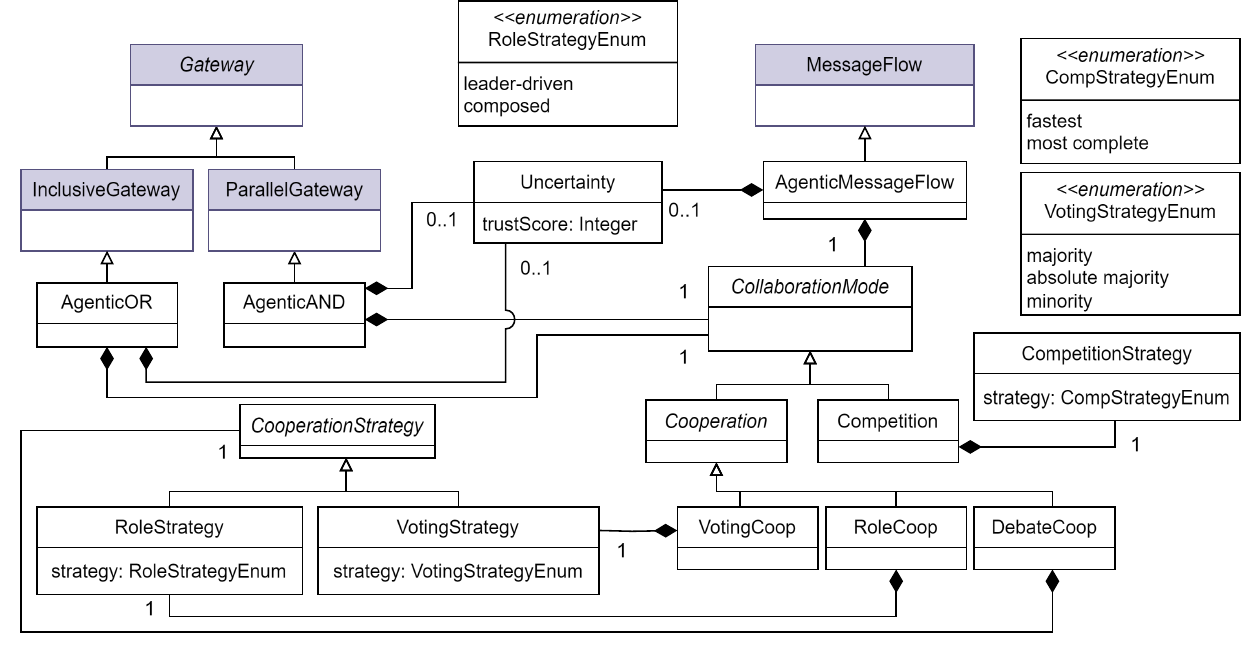}
    \caption{Domain model of the \textit{AgenticOR}, \textit{AgenticAND} and \textit{AgenticMessageFlow}.}
    \label{fig:model-agentic-gateways}
\end{figure}

In standard BPMN, collaboration between agents could be represented as groups, which allows for a basic depiction of collaborative efforts, but falls short in specifying the cooperation (and merging) strategies employed by the agents.

Figure~\ref{fig:model-agentic-gateways} shows the domain model of the collaboration types along with their merging strategies, and how they are related to the extended \textit{Gateway} and \textit{MessageFlow} classes.
The collaboration will be enclosed between diverging and merging gateways.
The former specifies the collaboration strategy, while the latter indicates the merging strategy.

We model the \textit{CollaborationMode} as the root of the hierarchy of collaboration modes, which can be either cooperation or competition, the former being the root of three types of cooperation (i.e., voting, role and debate). 
We extend both the inclusive and parallel gateways (see \textit{AgenticOR} and \textit{AgenticAND}, respectively), which also set the collaboration strategy (see \textit{CollaborationMode} association).
The inclusive gateway allows specifying a condition so that only specific agents are activated, while the parallel gateway diverges the flow to all its outputs, which is desired when all the participants from the diverged flow are required to collaborate.
Since agentic gateways are intended to represent collaboration, we only extend these two gateways.
We do not consider the exclusive gateway, since the flow is allowed only to one output.

The merging strategies set how to decide the solution to select in collaboration scenarios.
When competing, the \textit{CompetitionStrategy} class indicates the merging strategy. 
Regarding the cooperation strategies, the merging strategies are represented in the \textit{CooperationStrategy} class.
\textit{VotingStrategy} is used when the decision is made through voting, while \textit{RoleStrategy} is used when the merging is taken by a manager role or by the inherent roles of the agents (e.g., specialized agents proposing each part of the output).
While voting-based and role-based cooperation have explicit merging strategies because of their description, the debate-based cooperation might leverage from one of the two strategies.

The gateways allow the collaboration within pools, but if the set of agents is represented as another pool, message flows should be used (see \textit{MessageFlow}).
\textcolor{\mycolorrev}{We extend the message flow, rather than choreography tasks, since it allows for an elementary depiction of the collaboration.}
Like done for gateways, the outgoing message flow sets the collaboration strategy, while the incoming message flow sets the desired merging strategy.

Finally, the agentic gateways and message flow are also associated with a trust score.
In gateways, it can be used as a decision point, while in message flows can be used to understand the reliability of the received token.

\subsection{BPMN Notation Extension}
\label{sec:extension:notation}
According to the BPMN specification, extensions notation must not alter the notation of its elements, and must be as close as possible, in terms of look and feel, to it~\cite{BPMN_Specification_2014}.
Our extension introduces four new elements (see Table~\ref{tab:graphical-notation}) with a graphical representation close the BPMN element being extended. 

When choosing these elements, \textcolor{\mycolor}{we also aimed to stick to the principles for effective visual notations by Moody~\cite{moody2009physics}.} For instance, \textcolor{\mycolor}{to be compliant with the semiotic clarity principle, we maintain a 1:1 correspondence between semantic constructs and graphical symbols, with each agentic concept (lane, task, gateway, message flow) having its own distinct visual representation (see Table~\ref{tab:graphical-notation}).}

To identify the extended elements, we use a marker representing an agent.
\textcolor{\mycolor}{Our agent marker follows the perceptual discriminability principle, since it clearly distinguishes agentic elements from standard BPMN elements, while maintaining the basic shape of the original BPMN symbols to preserve familiarity.
Furthermore, the agent icon intuitively suggests intelligence and automation being semantically immediate, following the semantic transparency principle.}

\textcolor{\mycolor}{On the other hand, we minimize the introduction of new symbols by using a consistent agent marker with letter modifiers rather than creating entirely new shapes, following the graph economy and dual coding principle (i.e., use both graphical elements and textual annotations to enhance cognition).} 



The agentic lane is identified with the agent marker centered below the name for vertical lanes, or centered at the right of the name for horizontal lanes.
The trust score is set between the name of the lane and the agent marker, following the position of the text.
The role is set as a letter below the marker, where ``w'' stands for worker and ``m'' for manager.

The agentic task is represented with the agent marker in the top-left corner, as it is done with specific tasks (e.g., \textit{ManualTask}).
The reflection and collaboration strategies are indicated with a marker at the bottom of the shape.
The reflection strategy is set as a letter inside the marker, where ``s'' stands for self-reflection, ``c'' for cross-reflection, and ``h'' for human-reflection.
Note that the notation shown in Table~\ref{tab:graphical-notation} the  ``x'' is used as a placeholder.

The agentic gateways are represented with the agent marker at the top-left side, without disturbing the shape of the gateway.
The diverging gateway contains the collaboration marker in the bottom-right corner.
\textcolor{\mycolor}{This way the visual distance between symbols is greater~\cite{moody2009physics}.
The same principle has been applied for the remaining notations.}
The collaboration strategy is set as a letter below the marker, where ``c'' stands for competition, ``d'' for debate cooperation, ``r'' for role cooperation, and ``v'' for voting cooperation.
The merging gateway must contain the merging marker in the bottom-right corner.
The merging strategy is represented with two sets of letters below the marker, the first as the strategy class and the second as the merging strategy type.
Thus, for voting strategies are ``v-ma'' for majority, ``v-a'' for absolute majority, and ``v-mi'' for minority; for role-based strategies are ``r-l'' for leader-driven and ``r-c'' for composed; while for competition strategies are ``c-f'' for fastest, and ``c-mc'' for most complete. 

The agentic message flow is represented with the agent marker centered on the left side, if the message flow is vertical, or centered above the message flow, if it is horizontal.
The collaboration or merging strategy is set with a marker on the right side, if the message flow is vertical, or centered below the message flow, if it is horizontal.
As with the agentic gateway, the outgoing message flow must contain the collaboration strategy, while the incoming message flow must contain the merging strategy.

\begin{table}[t]
    \centering
    \caption{Graphical notation of the extended elements.}
    \label{tab:graphical-notation}
    \begin{tabularx}{.3\textwidth}{lc}
    \textsc{\makecell[l]{Extension element}} & \textsc{Notation} \\
    \toprule
    Agentic lane                             & \raisebox{-.5\height}{\includegraphics[height=15mm]{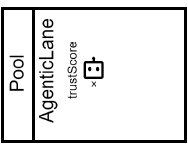}} \\
    \midrule
    Agentic task                             & \raisebox{-.5\height}{\includegraphics[height=10mm]{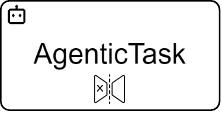}} \\
    \midrule
    \rule{0pt}{4ex}  
    \makecell[l]{Diverging agentic \\gateway} & \raisebox{-.5\height}{\includegraphics[height=10mm]{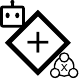}} \\
    \midrule
    \makecell[l]{Merging agentic \\gateway}   & \raisebox{-.5\height}{\includegraphics[height=10mm]{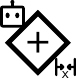}} \\ 
    \midrule
    \makecell[l]{Outgoing agentic\\ message flow} & \raisebox{-.5\height}{\includegraphics{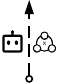}} \\
    \midrule
    \makecell[l]{Incoming agentic\\ message flow} & \raisebox{-.5\height}{\includegraphics{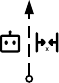}} \\
    \bottomrule
    \end{tabularx}
\end{table}

\subsection{Using our Extension to Model the Running Example}
\label{sec:extension:running-example}

\begin{figure*}[t]
    \centering
    \includegraphics[width=.8\textwidth]{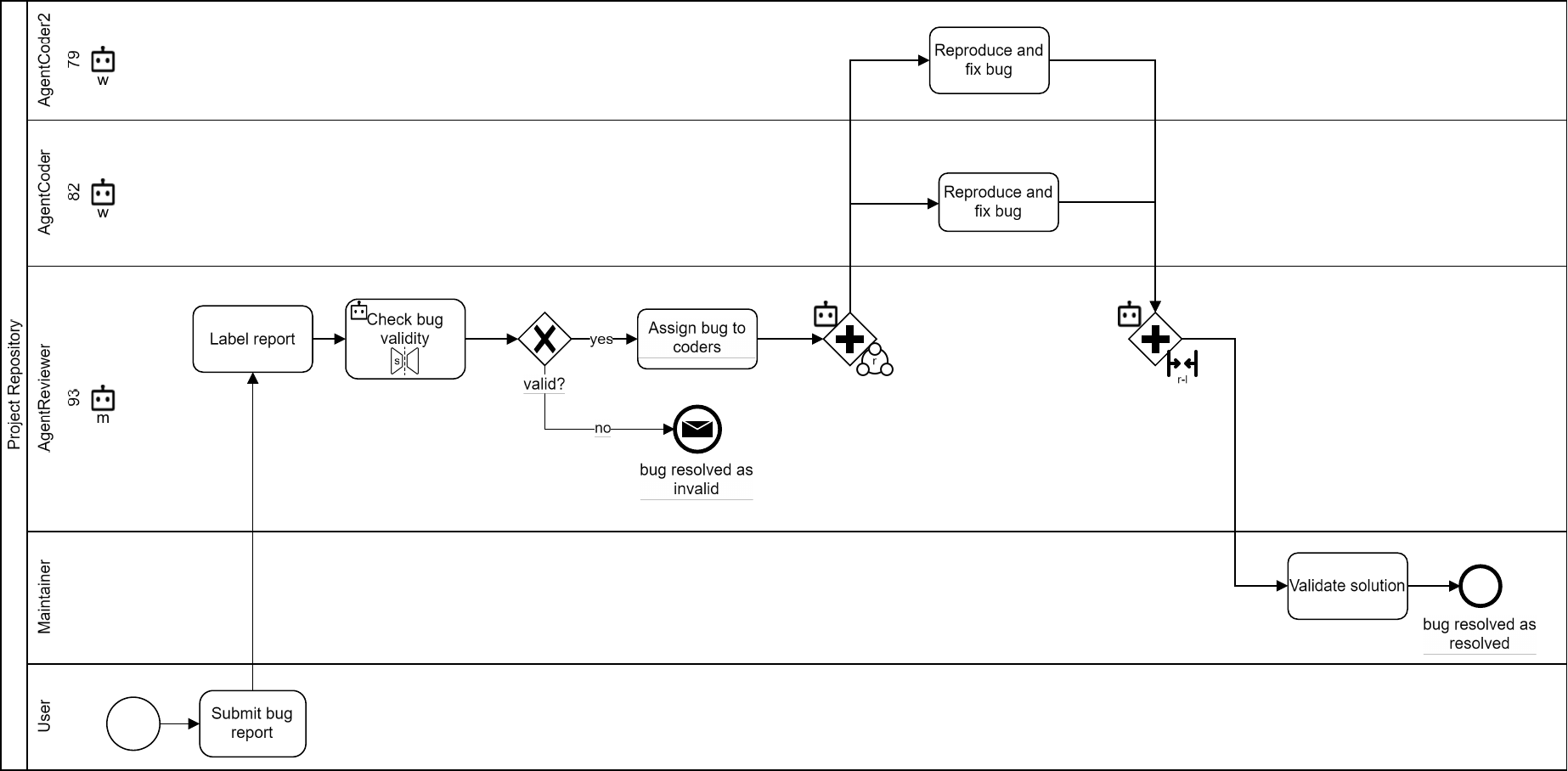}
    \caption{Running example with our extension notation.}
    \label{fig:running-example-extension}
\end{figure*}
Figure~\ref{fig:running-example-extension} shows the running example using our extension.
The different agents are denoted with \textit{agentic lanes}, and have a trust score attached (see three top lanes).
The second task (see \textit{Check bug validity}) is an \textit{agentic task} that applies self-reflection to the output.

To represent the collaboration between agents, we use the \textit{agentic gateway}, since agents are represented as lanes of the same pool.
The collaboration strategy applied is \textit{role cooperation}, as denoted by the notation.
After stating the collaboration strategy, the flow is divided towards the agents that collaborate.
All flows from this collaboration are merged into an \textit{agentic gateway} following the \textit{leader-driven}, strategy as denoted by the notation.
The remaining elements are compliant to standard BPMN.\looseness-1

\section{Our Extension as a Standard BPMN Extension Definition}
\label{sec:bpmn-x}

BPMN provides an extension mechanism to allow modeling domain-specific elements not included in the specification.

Nevertheless, there is a lack of methodological guides to develop and publish specific extensions. 
Stroppi et al.~\cite{DBLP:conf/bpmn/StroppiCV11} define a method (called BPMN+X) to transform a domain model into a BPMN-compliant extension by using UML profiles, which is the typical approach used in the UML world to define lightweight extensions\footnote{The term lightweight extension is used to denote language extensions that are compatible with language semantics and that can be expressed using the own language extension mechanisms, enabling the direct use of the extension in any tool that supports the language. This is in contrast to heavyweight extensions that offer more complex extensions that enable richer semantics for the extension but require dedicated tooling support} to the language. 
Given a BPMN extension defined as a profile, the BPMN+X method uses  mapping rules and automated model transformations to generate an XML Schema Extension Definition Document conforming to the official BPMN extension mechanism. 

\begin{figure*}[t]
    \centering
    \includegraphics[width=.9\textwidth]{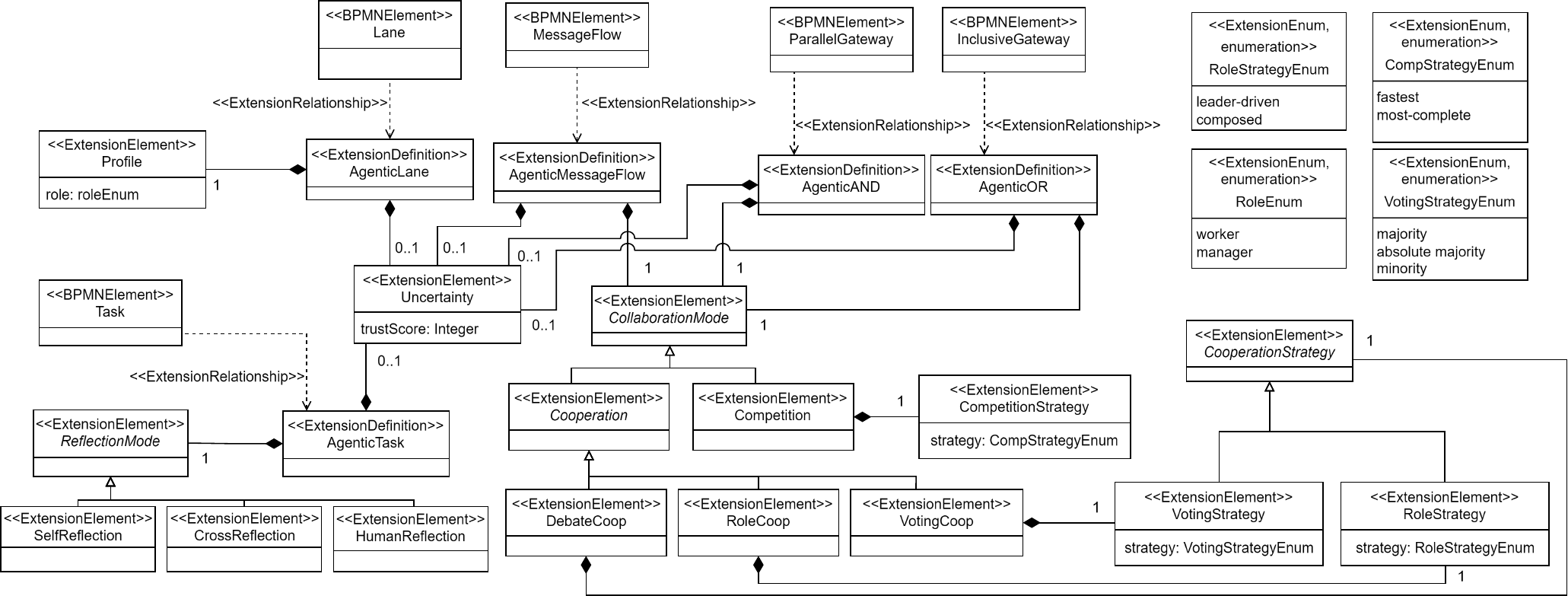}
    \caption{Extension model.}
    \label{fig:bpmn-extension-model}
\end{figure*}

We follow this approach to redefine our extended BPMN metamodel as a BPMN extension. To this purpose, Figure~\ref{fig:bpmn-extension-model} shows the metamodel from Figures~\ref{fig:model-agentic-lane}, \ref{fig:model-agentic-task} and \ref{fig:model-agentic-gateways} defined as a profile.\looseness-1

\section{Proof of Concept}
\label{sec:toolSupport}
As a proof-of-concept of the proposal, we have implemented a modeling editor that enables any developer to use our extended BPMN language and notation. 

\begin{figure}
    \centering
    \includegraphics[width=0.9\linewidth]{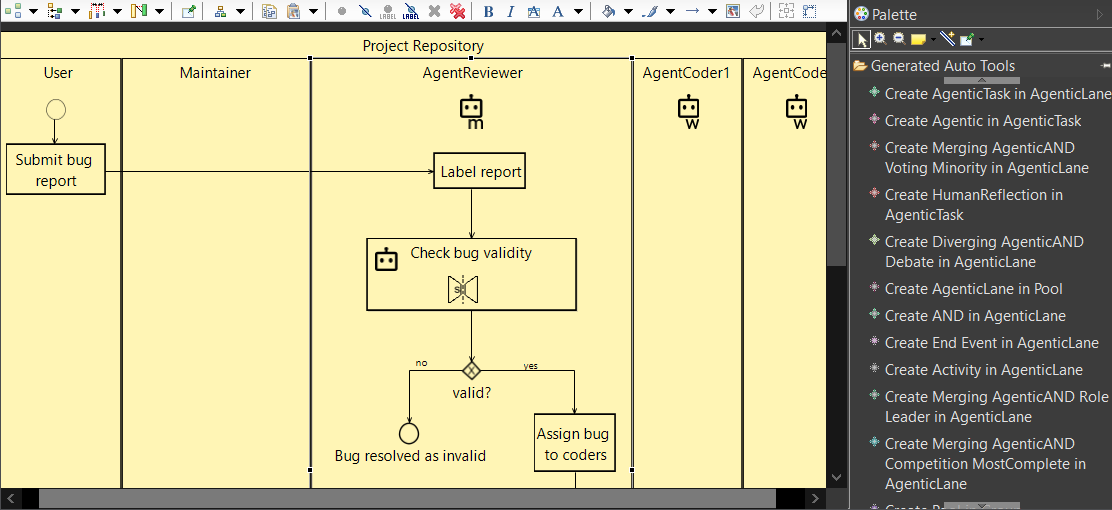}
    \caption{Platform-independent implementation.}
    \label{fig:example-aconite}
\end{figure}

The extension has been implemented using Sirius \footnote{\url{https://eclipse.dev/sirius/overview.html}}, an Eclipse project which allows you to easily create your own graphical modeling workbench by leveraging Eclipse Modeling technologies such as EMF and GMF.  Aconite~\cite{10.1145/3687997.3695642}, a tool that helps produce Sirius-based graphical notations, has been used to automatically generate the Sirius-based implementation. Figure~\ref{fig:example-aconite} shows a screenshot.

The tool repository\footnote{\url{https://github.com/BESSER-PEARL/agentic-bpmn}} includes the files that use the Aconite annotations and the examples illustrated in this paper.
The tool includes a main view of the diagram and a palette to allow the user to drag and drop the elements into the diagram. 
The palette contains the basic representation of BPMN, plus the extension elements proposed in this paper.
\textcolor{\mycolor}{Furthermore, the repository also includes several examples from the literature to illustrate the use of the extension.}

\section{Related Work}
\label{sec:related}
We classify the related work into two groups: proposals that try to model agents in general process modeling languages and agent-specific tools that may include a language to orchestrate agentic workflows.

In the first group we have K\"uster et al.~\cite{DBLP:journals/mags/KusterLHH12} that combine agent-oriented software engineering with business process design. Nevertheless, they do not extend BPMN and therefore suffer from the limitations stated in Section~\ref{sec:motivation}.  Endert et al.~\cite{endert2007mapping} propose a mapping of BDI agents to business processes. 
However, while this proposal provides ways of mapping agent-specific concepts following the BDI paradigm to BPMN, they do not cover advanced aspects of the current generation of agents such as reflection strategies.

Other BPMN approaches partially cover uncertainty concerns. 
Ceballos et al.~\cite{DBLP:conf/atal/CeballosFG15a} propose a BPMN Business Process Diagram (BPD) normal form based on Activity Theory~\cite{engestrom1999perspectives} that can be used for representing the dynamics of a collective human activity from the perspective of a subject.  
Herbert and Sharp~\cite{DBLP:journals/jcise/HerbertS13} proposed a BPMN extension introducing the uncertainty in sequence flows.
The extension includes probabilistic flows and rewards associated to the execution of tasks.
Note that the uncertainty is determined in the sequence flows rather than in the participant or activity BPMN element.
Our proposal introduces uncertainty at the participant level, then propagating to all the other elements. Both types of uncertainty could be combined.\looseness-1

Regarding specific tools to create agentic systems,  a couple include the graphical modeling of the process. Two representative examples are  
\textsc{LangGraph}\footnote{\url{https://www.langchain.com/langgraph}} that allows modeling the action flow of an agent (or a set of agents) using cyclic graphs and \textsc{Flowise}\footnote{\url{https://github.com/FlowiseAI/Flowise}}, an open-source low-code tool for orchestrating LLM agents.
In both cases, the workflows are purely focused on agent collaborations. Humans can only trigger the actions but are not supposed to collaborate with the agent to accomplish them. In contrast, our extension considers humans as active participants and enables representing complex human-agent interactions.  


\section{Conclusions and Further Work}
\label{sec:conclusion}
We have presented a BPMN extension to model human-agentic workflows.
Our extension enables the modeling of complex collaboration patterns between humans and agents, including specifying the agents' reflection strategies.  

As further work, we plan to provide a sublanguage to specify in detail the governance and decision-making strategies as part of the merging nodes for agents' collaboration efforts \textcolor{\mycolor}{(e.g., should decisions be based on consensus? voting strategies...).}
We will also propose an uncertainty propagation mechanism that, given the overall uncertainty of the agents and their confidence in the result of a given task, assigns an overall uncertainty of the task, which should then be propagated also to the consecutive tasks. 

Additionally, we plan to work on code generators aimed at producing an executable representation of the model, including the governance aspects and the uncertainty propagation mechanisms to make operational the modeled workflows. The generators could target, potentially,  a combination of BPEL engines and agentic platforms, where the BPEL engine would take care of the global orchestration and delegate to the agentic platform the task execution. 

Finally, we plan to empirically evaluate our extension and explore its application in industrial use cases.
\textcolor{\mycolor}{In particular, to ensure the cognitive effectiveness of our notation, we plan to conduct a metric-based assessment based on the alignment with Moody's principles for visual notations~\cite{moody2009physics}, following our previous work on collaborative modeling notation evaluation ~\cite{DBLP:conf/sle/0001CIM17}.}\looseness-1 


\bibliographystyle{splncs04}
\bibliography{bpmn-main}
\end{document}